\documentclass[aps,prc,twocolumn,10pt, superscriptaddress, showpacs, floatfix]{revtex4-1}

\usepackage{amssymb,epsfig}
\usepackage{bbold}

\newcommand{\be}{\begin{equation}}
\newcommand{\ee}{\end{equation}}
\newcommand{\bea}{\begin{eqnarray}}
\newcommand{\eea}{\end{eqnarray}}

\begin{document}

\title{A many-body approach to superfluid nuclei in axial geometry}
%\title{Toward a highly-accurate many-body theory: equation of motion method for the nuclear response}
%
\author{Yinu Zhang}
\affiliation{Department of Physics, Western Michigan University, Kalamazoo, MI 49008, USA}
\author{Antonio Bjel{\v{c}}i{\'c}}
\affiliation{Department of Physics, Faculty of Science, University of Zagreb, HR-10000 Zagreb, Croatia}
\author{Tamara Nik{\v{s}}i{\'c}}
\affiliation{Department of Physics, Faculty of Science, University of Zagreb, HR-10000 Zagreb, Croatia}
\author{Elena Litvinova}
\affiliation{Department of Physics, Western Michigan University, Kalamazoo, MI 49008, USA}
\affiliation{National Superconducting Cyclotron Laboratory, Michigan State University, East Lansing, MI 48824, USA}
\author{Peter Ring}
\affiliation{Fakult\"at f\"ur Physik, Technische Universit\"at M\"unchen, D-85748 Garching, Germany}
\author{Peter Schuck}
\affiliation{Institut de Physique Nucl\'eaire, IN2P3-CNRS, Universit\'e Paris-Sud, F-91406 Orsay Cedex, France}
\affiliation{Universit\'e Grenoble Alpes, CNRS, LPMMC, 38000 Grenoble, France}

%{Laboratoire de Physique et de Mod\'elisation des Milieux Condens\'es, CNRS et, Universit\'e Joseph Fourier, 25 Av. des Martyrs,
%BP 166, F-38042 Grenoble Cedex 9, France}

\date{\today}

\begin{abstract}
Starting from a general many-body fermionic Hamiltonian, we derive the equations of motion (EOM) for nucleonic propagators in a superfluid system. The resulting EOM is of the Dyson type formulated in the basis of Bogoliubov's quasiparticles. As the leading contributions to the dynamical kernel of this EOM in strongly-coupled regimes contain phonon degrees of freedom in various channels, an efficient method of calculating phonon's characteristics is required to successfully model these kernels. The traditional quasiparticle random phase approximation (QRPA) solvers are typically used for this purpose in nuclear structure calculations, however, they become very prohibitive in non-spherical geometries. In this work, by linking the notion of the quasiparticle-phonon vertex to the variation of the Bogoliubov's Hamiltonian, we show that the recently developed finite-amplitude method (FAM) can be efficiently employed to compute the vertices within the FAM-QRPA. To illustrate the validity of the method, calculations based on the relativistic density-dependent point-coupling Lagrangian are performed for the single-nucleon states in heavy and medium-mass nuclei with axial deformations. The cases of $^{38}$Si and $^{250}$Cf are presented and discussed.
\end{abstract}
%\pacs{21.10.-k, 21.30.Fe, 21.60.-n, 23.40.-s, 24.10.Cn, 24.30.Cz}

\maketitle

%===============================================================================
% Introduction
%===============================================================================
%{\it Introduction. \textemdash}
\section{Introduction}
Theoretical description of nuclear shell structure and response remain challenging aspects of nuclear physics for decades. The nuclear shell model pioneered by M. Goeppert-Mayer \cite{Goeppert-Mayer1975} and J.H.D. Jensen \cite{Jensen1949} and later promoted 
%by them 
to the inclusion of nuclear pairing \cite{Mayer1950,50BCS} has provided the essential building blocks for understanding the fermionic motion in medium-mass and heavy nuclei. The paradigm of the mean field dominating higher-rank fermionic correlations was developed throughout further decades into the sophisticated microscopic self-consistent mean fields linked to the density functional theory (DFT) 
\cite{Bender2003,VretenarAfanasjevLalazissisEtAl2005,Stoitsov2009,Afanasjev2013,Bogner2013,Meng2016,Reinhard2018,Colo2020}, 
which are capable of reproducing the experimentally established nuclear shells, both spherical and deformed, reasonably well.

With the advent of the radioactive beam facilities the concept of firm nuclear shells and well-defined magic numbers associated with the enhanced stability of closed-shell nuclei started to change. It turned out, in particular, that the unstable systems with exotic neutron-to-proton ratios may exhibit magic numbers, which are different from those in stable nuclei. This phenomenon is studied extensively, both experimentally and theoretically, and there are indications that it can be associated with the enhanced role of beyond-mean-field correlations in exotic nuclear systems \cite{Baumann2007,Gade2008,Jensen2011,Otsuka2020,Aumann2021}. Although the criteria of magicity are not unambiguously defined and can be associated with the shell gaps, the peculiarities in the systematic behavior of the lowest quadrupole states or charge radii, the idea of violation of magic numbers in nuclei with extreme neutron-to-proton ratios is widely accepted \cite{Otsuka2020,Aumann2021}. 

%Deformed shell closures
%medium-mass 
%heavy and superheavy
%underlying nuclear many-body problem
%potential to be linked to the optical potential and reaction theory

As  many of the successful density functionals are based on a considerably reduced effective nucleon mass, as compared to its bare values, they typically underestimate the fermionic level densities and overestimate the respective occupation probabilities 
\cite{Afanasjev2003,Afanasjev2011,Dobaczewski2015}. The inclusion of correlations beyond the mean field helps resolving these deficiencies and can be done by taking into account the dynamical part of the nucleonic self-energy, which arises from the model-independent equations of motion (EOM) for the in-medium fermionic propagator \cite{LitvinovaSchuck2019,Schuck2021} and which is neglected in the DFT. This leads to the fragmentation of the mean-field states and the densifying of the single-particle spectra \cite{RingWerner1973,BertschBortignonBroglia1983,LitvinovaRing2006,LitvinovaAfanasjev2011,AfanasjevLitvinova2015,Litvinova2012,Litvinova2016}. 

The important ingredients for the dynamical self-energy in the leading approximation are the particle-vibration coupling (PVC) vertices and the frequencies of the vibrational modes (phonons). In the DFT-based self-consistent approaches they can be calculated within the (quasiparticle) random phase approximation ((Q)RPA). 
This strategy based on the traditional QRPA diagonalization solvers works reasonably well for spherically-symmetric nuclear systems, however, it becomes very prohibitive for calculations in non-spherical geometries.  This fact limited the existing applications of the DFT-PVC method to only spherical nuclei.

In this work, we report the first results of the approach designed to overcome this limitation.
We employ the recently developed finite-amplitude method (FAM) to solve the relativistic QRPA equations in the deformed Dirac-Hartree-Bogoliubov basis for the variations of the fermionic density \cite{Bjelcic2020} and extract the PVC vertices  by linking these solutions to the fermionic dynamical self-energy obtained within the EOM method \cite{LitvinovaSchuck2019} generalized for the superfluid phase.
As the FAM has manifested itself over the past decade as a very efficient method for numerical solutions of the RPA and QRPA  equations \cite{Nakatsukasa2007,Kortelainen2015,Niksic2013,Bjelcic2020}, we, thereby, utilize the advantages of the FAM for extending the mean-field theory for non-spherical systems and present the first numerical implementations of this extension for nuclei with axial deformations.

%===============================================================================
% Formalism
%===============================================================================

%{\it Formalism. \textemdash}
\section{Formalism}

\subsection{Equation of motion for the quasiparticle propagator}

In this work we consider the equation of motion for the quasiparticle propagator derived ab initio, i.e., with the only input of the bare nucleon-nucleon interaction in the vacuum. As it was discussed in Refs. \cite{LitvinovaSchuck2019,Schuck2021,Litvinova2021}, such an approach allows one to obtain the most general and model-independent expressions for the EOM interaction kernels, which can be then approximated with various degrees of accuracy and adopted for calculations with effective interactions. 
Thus, the starting point is the many-body Hamiltonian
 \be
H = H^{(1)} + V = \sum_{12}h_{12}\psi^{\dag}_1\psi_2 + \frac{1}{4}\sum\limits_{1234}{\bar v}_{1234}{\psi^{\dagger}}_1{\psi^{\dagger}}_2\psi_4\psi_3
\label{Hamiltonian}
\ee  
with the one-body matrix elements $h_{12}$ comprising the kinetic energy and the external mean field in case it is present, and 
the antisymmetrized matrix elements of the two body-interaction ${\bar v}_{1234}$. The three-body forces are neglected in the present study, but can be included as an extension of the framework. The number indices stand for complete sets of quantum numbers defining the single-particle degrees of freedom. Furthermore, it is convenient to work in the canonical basis, which diagonalizes the one-body part of the Hamiltonian, so that we set $h_{12} = \delta_{12}\varepsilon_{1}$.

\begin{figure*}
\begin{center}
\vspace{0.3cm}
\includegraphics[scale=0.8]{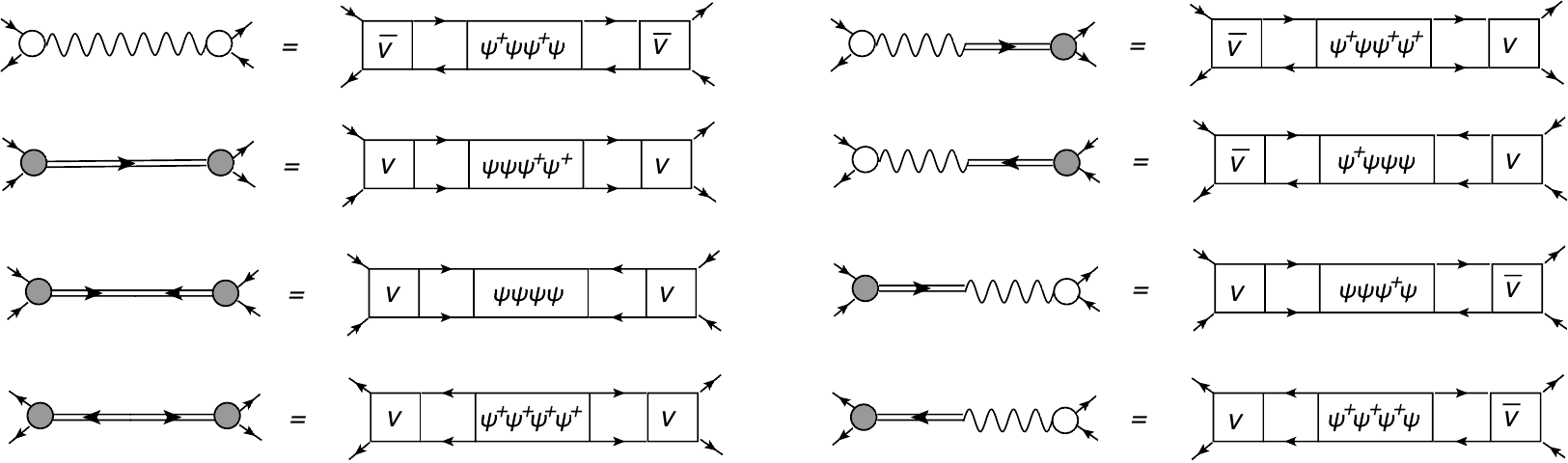}
\end{center}
\caption{The emergent origin of the quasiparticle-vibration coupling amplitudes in the diagrammatic form. Empty and filled circles denote normal and pairing vibration (phonon) vertices, while the wavy and double lines with arrows are the normal and pairing phonon propagators, respectively. The square boxes stand for the antisymmetrized $\bar v$ and non-antisymmetrized $v$ interaction matrix elements. The operator products in the rectangular boxes, together with the attached fermionic lines (solid lines with arrows), are reserved for the two-point correlation functions, with the following correspondence: \framebox{abcd} $= -i\langle T(ab)(t)(cd)(t')\rangle$.}
\label{QVC_map}
\end{figure*}

The quasiparticle propagator ${\hat G}_{12}$ through the superfluid correlated medium is defined as follows:
\bea
{\hat G}_{12}(t-t') = -i\left( \begin{array}{cc} \langle T\psi_1(t)\psi^{\dagger}_2(t')\rangle &  \langle T\psi_1(t)\psi_2(t')\rangle \\
 \langle T\psi^{\dagger}_1(t)\psi^{\dagger}_2(t')\rangle &  \langle T\psi^{\dagger}_1(t)\psi_2(t')\rangle
\end{array} \right)  \nonumber\\
\equiv \left( \begin{array}{cc} G^{(11)}_{12}(\tau)  & G^{(12)}_{12}(\tau) \\ 
G^{(21)}_{12}(\tau) & G^{(22)}_{12}(\tau)\end{array}\right) \equiv 
\left( \begin{array}{cc} G_{12}(\tau)  &  F^{(1)}_{12}(\tau) \\ 
F^{(2)}_{12}(\tau) & G^{(h)}_{12}(\tau)\end{array}\right)\nonumber\\
\label{Gq}
\eea
with $T$ being the time ordering operator, $\tau = t-t'$, and $\psi_1(t), \psi^{\dagger}_1(t)$ the time-dependent fermionic field operators in Heisenberg representation. The propagator (\ref{Gq}) is often called Gor'kov propagator \cite{Abrikosov1965} and famously includes both normal components on the main diagonal and anomalous off-diagonal components, which are compatible with the relaxed particle number conservation condition. 
The most direct way for generating a time-dependent EOM for this propagator is differentiation with respect to the time variables.  Differentiating with respect to $t$, then with respect to $t'$ and performing the Fourier transformation with respect to $\tau$ to the domain of the energy variable $\varepsilon$ \cite{LitvinovaSchuck2019,Schuck2021,Litvinova2021} leads to the following equation:
\bea
\left(\begin{array}{cc}G_{11'}(\varepsilon) & F^{(1)}_{11'}(\varepsilon) \\ F^{(2)}_{11'}(\varepsilon) & G^{(h)}_{11'}(\varepsilon)\end{array}\right) = \left(\begin{array}{cc}G^0_{11'}(\varepsilon) & 0 \\ 0 & G^{(h)0}_{11'}(\varepsilon)\end{array}\right) \nonumber\\
+ 
\sum\limits_{22'}\left(\begin{array}{cc}G^0_{12}(\varepsilon) & 0 \\ 0 & G^{(h)0}_{12}(\varepsilon)\end{array}\right)\left(\begin{array}{cc}T_{22'}(\varepsilon) & T^{(1)}_{22'}(\varepsilon) \\ T^{(2)}_{22'}(\varepsilon) & T^{(h)}_{22'}(\varepsilon) \end{array}\right)\nonumber\\ \times
\left(\begin{array}{cc}G^0_{2'1'}(\varepsilon) & 0 \\ 0 & G^{(h)0}_{2'1'}(\varepsilon)\end{array}\right)\nonumber\\
\label{GTgen0}
\eea
or, symbolically,
\be
{\hat G}_{11'}(\varepsilon) = {\hat G}^{0}_{11'}(\varepsilon) + \sum\limits_{22'}{\hat G}^0_{12}(\varepsilon){\hat T}_{22'}(\varepsilon){\hat G}^0_{2'1'}
(\varepsilon).
\label{GTgen}
\ee
In Eq. (\ref{GTgen}) ${\hat G}^{0}$ is the free quasiparticle propagator 
\bea
{\hat G}^{0}_{11'}(\varepsilon) = \left(\begin{array}{cc}G^0_{11'}(\varepsilon) & 0 \\ 0 & G^{(h)0}_{11'}(\varepsilon)\end{array}\right) \nonumber\\ 
=
\left(\begin{array}{cc} {\delta_{11'}}/{(\varepsilon - \varepsilon_1)} & 0 \\ 0 & {\delta_{11'}}/{(\varepsilon+ \varepsilon_1)} \end{array}\right), 
\label{Gfree}
\eea
with vanishing anomalous components,
and ${\hat T}$ is the quasiparticle $T$-matrix of the following origin:
 \bea
 {\hat T}_{11'}(t-t')  =  {\hat T}^{0}_{11'}(t-t') +  {\hat T}^r_{11'}(t-t')  \nonumber \\
 = -\delta(t-t')\langle \left(\begin{array}{cc}  \bigl[ [V,\psi_1],{\psi^{\dagger}}_{1'}\bigr]_+ & [[V,\psi_1],\psi_{1'}]_+ \\ 
\bigl[ [V,\psi^{\dagger}_1],{\psi^{\dagger}}_{1'}\bigr]_+ & \bigl[ [V,\psi^{\dagger}_1],{\psi}_{1'}\bigr]_+
 \end{array}\right)\rangle \nonumber \\ + 
i\langle T\left(\begin{array}{cc}  \bigl[V,\psi_1\bigr](t)\bigl[V,\psi^{\dagger}_{1'}\bigr](t') & \bigl [V,\psi_1\bigr](t)\bigl[V,\psi_{1'}\bigr](t') \\  
\bigl[V,\psi^{\dagger}_1\bigr](t)\bigl[V,\psi^{\dagger}_{1'}\bigr](t') &  \bigl[V,\psi^{\dagger}_1\bigr](t)\bigl[V,\psi_{1'}\bigr](t') \end{array}\right)\rangle ,
\nonumber\\
\label{Tmatrix}
 \eea
where we adopted the notation $(AB)(t) = e^{iHt}ABe^{-iHt}$ for the Heisenberg representation of the operator products and an analogous convention for the commutators. 
The matrix ${\hat T}_{11'}(\varepsilon)$ is the Fourier image of $ {\hat T}_{11'}(t-t')$ in the energy domain. The important feature of the $T$-matrix (\ref{Tmatrix}) is its decomposition into the static (instantaneous) ${\hat T}^{0}$ and dynamical ${\hat T}^{r}$ components. The static component is independent of time (energy) and reads:
\bea
{\hat T}^0_{11'} = \sum\limits_{ij}\left(\begin{array}{cc}  {\bar v}_{1i1'j}\langle \psi^{\dagger}_i\psi_j\rangle & \frac{1}{2}{\bar v}_{11'ij}\langle \psi_j\psi_i\rangle
\\  \frac{1}{2}{\bar v}_{ij11'}\langle \psi^{\dagger}_j\psi^{\dagger}_i\rangle & -{\bar v}_{1'i1j}\langle \psi^{\dagger}_i\psi_j\rangle
\end{array}\right) \nonumber \\ \equiv
\left(\begin{array}{cc} {\tilde\Sigma}_{11'} & \Delta_{11'} \\ -\Delta^{\ast}_{11'} & -{\tilde\Sigma}^{T}_{11'}\end{array}\right),
\label{T0}
\eea
comprising the single-particle and single-hole mean fields on the main diagonal as well as the off-diagonal pairing fields. 
The dynamical component in the time domain is also a 2$\times$2 array
\be
{\hat T}^r_{11'}(t-t') = \left(\begin{array}{cc} {\hat T}^{r}_{11'}(t-t') & {\hat T}^{r(1)}_{11'}(t-t')\\
{\hat T}^{r(2)}_{11'}(t-t') & {\hat T}^{r(h)}_{11'}(t-t') \end{array}\right),
\ee
where
\bea
{\hat T}^r_{11'}(\tau) = \frac{-i}{4}\sum\limits_{iklpqr}
{\bar v}_{1ikl}\langle T(\psi^{\dagger}_i\psi_l\psi_k)(t)(\psi^{\dagger}_q\psi^{\dagger}_p\psi_r)(t')\rangle {\bar v}_{pqr1'} \nonumber\\
{\hat T}^{r(1)}_{11'}(\tau) = \frac{-i}{4}\sum\limits_{iklpqr}
{\bar v}_{1ikl}\langle T(\psi^{\dagger}_i\psi_l\psi_k)(t)(\psi^{\dagger}_p\psi_r\psi_q)(t')\rangle {\bar v}_{p1'qr} \nonumber\\
{\hat T}^{r(2)}_{11'}(\tau) = \frac{-i}{4}\sum\limits_{iklpqr}
{\bar v}_{ikl1}\langle T(\psi^{\dagger}_i\psi^{\dagger}_k\psi_l)(t)(\psi^{\dagger}_q\psi^{\dagger}_p\psi_r)(t')\rangle {\bar v}_{pqr1'} \nonumber\\
{\hat T}^{r(h)}_{11'}(\tau) = \frac{-i}{4}\sum\limits_{iklpqr}
{\bar v}_{ikl1}\langle T(\psi^{\dagger}_i\psi^{\dagger}_k\psi_l)(t)(\psi^{\dagger}_p\psi_r\psi_q)(t')\rangle {\bar v}_{p1'qr}, \nonumber\\
\label{Tr} 
\eea
and consists of the double convolutions of three-fermion two-point propagators with the interaction matrix elements. Introducing the irreducible with respect to the free propagator ${\hat G}^{0}_{11'}$ (\ref{Gfree}) part of the $T$-matrix ${\hat\Sigma} = {\hat T}^{irr}$, Eq. (\ref{GTgen}) can be written in the Dyson form:
\be
{\hat G}_{11'}(\varepsilon) = {\hat G}^{0}_{11'}(\varepsilon) + \sum\limits_{22'}{\hat G}^0_{12}(\varepsilon){\hat \Sigma}_{22'}(\varepsilon){\hat G}_{2'1'}
(\varepsilon),
\label{GDysongen}
\ee
where the self-energy is obviously decomposed into the static ${\hat\Sigma}^0$ and dynamical ${\hat\Sigma}^r$ parts: ${\hat\Sigma} = {\hat\Sigma}^0 + {\hat\Sigma}^r$ with ${\hat\Sigma}^0 = {\hat T}^0$ and ${\hat\Sigma}^r = {\hat T}^{r;irr}$. The static, or mean-field, contribution is determined by Eq. (\ref{T0}) through the normal $\langle \psi^{\dagger}_i\psi_j\rangle$ and pairing $\langle \psi_i\psi_j\rangle$ one-body densities, which can, in principle, be found self-consistently as the static limit of $\hat G$. The dynamical contribution is the irreducible part of Eq. (\ref{Tr}), which contains three-fermion propagators. These propagators can be with minimal approximations decomposed into the products of one-fermion and two-fermion propagators as described in detail in Refs. \cite{LitvinovaSchuck2019,Litvinova2021}. Here, as in 
Ref. \cite{Litvinova2021}, we retain all the possible irreducible combinations including those with anomalous one-body and two-body propagators. The two-body propagators enter the dynamical self-energy components as double contractions with the interaction matrix elements. The complete set of those combinations is shown diagrammatically in Fig.  \ref{QVC_map}, where it is mapped onto the set of quasiparticle-vibration coupling (qPVC) amplitudes. Here and hereinafter by the abbreviation "qPVC" we emphasize that the PVC is taken into account consistently and on equal footing with superfluidity.

The mapping of Fig. \ref{QVC_map} is exact and independent of the approximation, which is made for the two-body propagators. In practice, the propagators can be found from the EOMs generated for each of them, as it is discussed in Refs. \cite{LitvinovaSchuck2019,LitvinovaSchuck2020}. In a fully ab-initio approach, these EOMs should be solved self-consistently together with the quasiparticle propagator (\ref{Gq}), in a certain approximation for the two-body dynamical kernels. While such a realization will be performed elsewhere, in the present work the static kernel is taken in the form of the effective interaction as the second variational derivative of the energy density functional (EDF) with respect to the superfluid density matrix. The phonon vertices, respectively, are computed with this interaction. If the EDF is adjusted to finite nuclei, the QRPA provides a good first approximation to the phonon characteristics. 

The EDFs also represent an appropriate starting point to describe the quasiparticle states. Therefore, it is convenient to recast the Dyson Eq. (\ref{GDysongen}) in terms of the mean-field propagator ${\hat{\tilde G}}$:
\be
{\hat G}_{11'}(\varepsilon) = {\hat{\tilde G}}_{11'}(\varepsilon) + \sum\limits_{22'}{\hat{\tilde G}}_{12}(\varepsilon){\hat \Sigma}^r_{22'}(\varepsilon){\hat G}_{2'1'}
(\varepsilon), 
\label{GDysongen1}
\ee
such that
\be
{\hat{\tilde G}}_{11'}(\varepsilon) = {\hat G}^{0}_{11'}(\varepsilon) + \sum\limits_{22'}{\hat G}^0_{12}(\varepsilon){\hat \Sigma}^0_{22'}(\varepsilon){\hat{\tilde G}}_{2'1'}
(\varepsilon),
\label{GDysongenMF}
\ee
which is the direct output of the EDF. 

\begin{figure*}
\begin{center}
\includegraphics[scale=0.33]{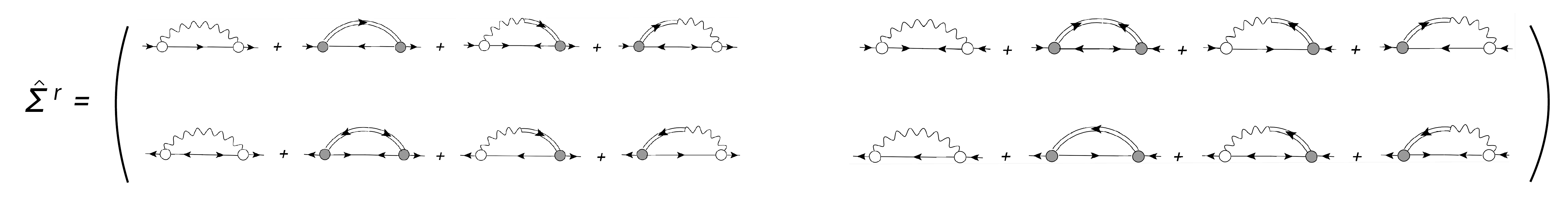}
\end{center}
\caption{ The $2\times2$ matrix structure of the dynamical self-energy in the canonical basis. The diagrammatic conventions for the phonon propagators, single-arrow fermionic lines and PVC vertices are the same as in Fig. \ref{QVC_map}. The single fermionic lines with double arrows denote the anomalous one-fermion propagators: right-left arrows are associated with $F^{(1)}$, and left-right arrows stand for $F^{(2)}$.}
\label{SEr}
\end{figure*}

The transformation of ${\hat{\tilde G}}$ and ${\hat G}$ to the quasiparticle basis $|\nu\rangle$ singles out their forward (+) and backward (-) components
\bea
{\tilde G}^{(+)}_{\nu\nu'}(\varepsilon) = \sum\limits_{12} \Bigl(U^{\dagger}_{\nu 1} \ \ \  V^{\dagger}_{\nu 1} \Bigr) {\hat {\tilde G}}_{12}(\varepsilon) 
\left( \begin{array}{c} U_{2\nu'} \\ V_{2\nu'} \end{array}\right) \nonumber\\
{G}^{(+)}_{\nu\nu'}(\varepsilon) = \sum\limits_{12} \Bigl(U^{\dagger}_{\nu 1} \ \ \  V^{\dagger}_{\nu 1} \Bigr) {\hat {G}}_{12}(\varepsilon) 
\left( \begin{array}{c} U_{2\nu'} \\ V_{2\nu'} \end{array}\right) 
%\label{UV+}
\nonumber\\
{\tilde G}^{(-)}_{\nu\nu'}(\varepsilon) = \sum\limits_{12} \Bigl(V^{T}_{\nu 1} \ \ \  U^{T}_{\nu 1} \Bigr) {\hat {\tilde G}}_{12}(\varepsilon) 
\left( \begin{array}{c} V^{\ast}_{2\nu'} \\ U^{\ast}_{2\nu'} \end{array}\right) \nonumber\\
{G}^{(-)}_{\nu\nu'}(\varepsilon) = \sum\limits_{12} \Bigl(V^{T}_{\nu 1} \ \ \  U^{T}_{\nu 1} \Bigr) {\hat {G}}_{12}(\varepsilon) 
\left( \begin{array}{c} V^{\ast}_{2\nu'} \\ U^{\ast}_{2\nu'} \end{array}\right),
\label{UV+-}
\eea
with the aid of the Bogoliubov's matrices $U_{1\nu}$ and $V_{1\nu}$ connecting the particle $\psi_1$ and quasiparticle $\alpha_{\nu}$ Fock operators:
\be
\psi_1 = \sum\limits_{\nu} \bigl(U_{1\nu}\alpha_{\nu} + V^{\ast}_{1\nu}\alpha^{\dagger}_{\nu}\bigr),\ \ \ \ \ \ \ \ \ \ \ 
\psi^{\dagger}_1 = \sum\limits_{\nu} \bigl(V_{1\nu}\alpha_{\nu} + U^{\ast}_{1\nu}\alpha^{\dagger}_{\nu}\bigr).
\label{Btrans}
\ee
Applying the transformations (\ref{UV+-}) to the Dyson equation (\ref{GDysongen1}) leads to:
\be
G^{(\eta)}_{\nu\nu'}(\varepsilon) = {\tilde G}^{(\eta)}_{\nu\nu'}(\varepsilon) + \sum\limits_{\mu\mu'}{\tilde G}^{(\eta)}_{\nu\mu}(\varepsilon)\Sigma^{r(\eta)}_{\mu\mu'}(\varepsilon)G^{(\eta)}_{\mu'\nu'}(\varepsilon), 
\label{Dyson_qp}
\ee
where $(\eta) = (+)$ and $(\eta) = (-)$. The components of the 
dynamical kernel are transformed to the quasiparticle space, accordingly, as
\bea
\Sigma^{r(+)}_{\mu\mu'}(\varepsilon) = \sum\limits_{12} \Bigl(U^{\dagger}_{\mu 1} \ \ \  V^{\dagger}_{\mu 1} \Bigr) {\hat\Sigma}^r_{12}(\varepsilon)
\left( \begin{array}{c} U_{2\mu'} \\ V_{2\mu'} \end{array}\right)  \nonumber\\
\Sigma^{r(-)}_{\mu\mu'}(\varepsilon) = \sum\limits_{12} \Bigl(V^{T}_{\mu 1} \ \ \  U^{T}_{\mu 1} \Bigr) {\hat\Sigma}^r_{12}(\varepsilon)
\left( \begin{array}{c} V^{\ast}_{2\mu'} \\ U^{\ast}_{2\mu'} \end{array}\right), 
\label{Sigma_pm}
\eea
while the explicit form of the dynamical self-energy in the canonical basis ${\hat\Sigma}^r_{12}$ is represented diagrammatically in Fig. \ref{SEr}. It contains  all possible convolutions of the amplitudes listed in Fig. \ref{QVC_map} with the normal and anomalous one-fermion propagators (\ref{Gq}). The corresponding analytical derivation for ${\hat\Sigma}^r_{12}$ can be found in Ref. \cite{Litvinova2021}. 
The mean-field and the exact quasiparticle propagators, respectively, read:
\bea
{\tilde G}^{(\eta)}_{\nu\nu'}(\varepsilon) = \frac{\delta_{\nu\nu'}}{\varepsilon - \eta(E_{\nu} - E_0 - i\delta)}, \nonumber\\
{G}^{(\eta)}_{\nu\nu'}(\varepsilon) = \sum\limits_n\frac{S^{(\eta)n}_{\nu\nu'}}{\varepsilon - \eta({\cal E}_{n} - E_0 - i\delta)},
\label{G_qp}
\eea
where the summation is formally running over the complete set of states $|n\rangle $ in $(N+1)$-particle system for $(\eta) = (+)$ and in $(N-1)$-particle system for $(\eta) = (-)$. 
One can see that after the transformations (\ref{UV+-}) to the quasiparticle basis the solution of the Dyson equation in the form of Eq. (\ref{Dyson_qp}) reduces to finding only two components of the quasiparticle propagator, instead of four of them in the canonical basis of Eq. (\ref{Gq}). The computational effort reduces considerably as the equations for the $(\eta) = (\pm)$ components are decoupled.
Furthermore, the residues in the exact propagator of Eq. (\ref{G_qp}), also called spectroscopic factors, coincide, i.e., $S^{(+)n}_{\nu\nu'} = S^{(-)n}_{\nu\nu'}$ if the particle number conservation condition is relaxed, that is the case in the Bogoliubov's theory and in QRPA. %
This means that only one of the Eqs. (\ref{Dyson_qp}) needs to be solved, e.g., the one for $(\eta) = (+)$, that further reduces the computation effort by a factor of two. 

The remaining quantity to be determined is, thus, the dynamical self-energy in the quasiparticle basis.  As a result of the transformation (\ref{Sigma_pm}), its forward component $(\eta) = (+)$ reads:
\bea
\Sigma^{r(+)}_{\nu\nu'}(\varepsilon) = \sum\limits_{\nu''\mu} \Bigl[ 
\frac{{\Gamma}^{(11)\mu}_{\nu\nu''}{\Gamma}^{(11)\mu\ast}_{\nu'\nu''}}{\varepsilon  - E_{\nu''} - \omega_{\mu} + i\delta} +
\frac{{\Gamma}^{(02)\mu\ast}_{\nu\nu''}{\Gamma}^{(02)\mu}_{\nu'\nu''}}{\varepsilon + E_{\nu''} + \omega_{\mu} - i\delta} \Bigl]\nonumber\\ 
\label{SEqp}
\eea
where $\omega_{\mu}$ are the frequencies of the superfluid phonons, which combine normal and pairing phonons, as shown diagrammatically in Fig. \ref{SEr}.
The corresponding superfluid phonon vertices  (the qPVC vertices) $\Gamma^{(11)\mu}$ and $\Gamma^{(02)\mu}$ contain, respectively, the following linear combinations of the normal $g^{\mu}$ and pairing $\gamma^{\mu(\pm)}$ phonon vertices:
\bea
\Gamma^{(11)\mu}_{\nu\nu'} = \sum\limits_{12}\Bigl[ 
U^{\dagger}_{\nu 1}g^{\mu}_{12}U_{2\nu'} + U^{\dagger}_{\nu 1}\gamma^{\mu(+)}_{12}V_{2\nu'} \nonumber\\
- V^{\dagger}_{\nu 1}(g^{\mu }_{12})^TV_{2\nu'} - V^{\dagger}_{\nu 1}(\gamma^{\mu(-)}_{12})^TU_{2\nu'}\Bigr]  
%\equiv 
%\Bigl[ 
%U^{\dagger}g^{\mu}U + U^{\dagger}\gamma^{\mu(+)}V 
%- V^{\dagger}g^{\mu T}V - V^{\dagger}\gamma^{\mu(-)T}U\Bigr]_{\nu\nu'} 
\label{Gamma11_HFB}
\\
\Gamma^{(02)\mu}_{\nu\nu'} = -\sum\limits_{12}\Bigl[ 
V^{T}_{\nu 1}g^{\mu}_{12}U_{2\nu'} + V^{T}_{\nu 1}\gamma^{\mu(+)}_{12}V_{2\nu'} \nonumber\\
- U^{T}_{\nu 1}(g^{\mu }_{12})^TV_{2\nu'} - U^{T}_{\nu 1}(\gamma^{\mu(-)}_{12})^TU_{2\nu'}\Bigr], 
%\equiv 
%-\Bigl[ 
%V^{T}g^{\mu}U+ V^{T}\gamma^{\mu(+)}V 
%- U^{T}g^{\mu T}V - U^{T}\gamma^{\mu(-)T}U\Bigr]_{\nu\nu'}.
\label{Gamma02_HFB}
\eea
defined as:
\bea
g^{\mu}_{13} &=& \sum\limits_{24}{\bar v}_{1234}\rho^{\mu}_{42} \nonumber\\
\gamma^{\mu(+)}_{12} &=& \sum\limits_{34} v_{1234}\varkappa^{\mu(+)}_{34}, \ \ \ \ \  \gamma_{12}^{\mu(-)} = 
\sum\limits_{34} \varkappa^{\mu(-)\ast}_{34} v_{3412}, \nonumber\\
\eea
via the normal $\rho^{\mu}$ and pairing $\varkappa^{\mu(\pm)}$ transition densities
\bea
\rho^{\mu}_{42} = \langle 0|\psi^{\dagger}_2\psi_4|\mu \rangle \nonumber\\
\varkappa^{\mu(+)}_{34} = \langle 0 | \psi_4\psi_3|\mu \rangle,
\ \ \ \ \  \varkappa^{\mu(-)\ast}_{34} = \langle 0 | \psi^{\dagger}_4\psi^{\dagger}_3|\mu \rangle, \nonumber\\
\label{trden}
\eea
where $|0\rangle$ and $|\mu\rangle$ are the ground and excited states of the even-even core, respectively.
Eqs. (\ref{Gamma11_HFB},\ref{Gamma02_HFB}) are obtained under the assumptions that the intermediate quasiparticle propagators in the dynamical self-energy given by Fig. \ref{SEr} are the superfluid mean-field (Hartree-Fock-Bogoliubov (HFB) or Hartree-Bogoliubov) propagators \cite{Litvinova2021}. This is a rather good approximation within EDF frameworks, while further more accurate solutions can be obtained by iterating these propagators in a self-consistent cycle. 

\subsection{Phonon vertex extraction from FAM-QRPA} 

Determining the phonon characteristics is an external procedure with respect to the Dyson equation and requires, in general, solving an EOM for the superfluid two-fermion propagator, or response function. Although quite advanced solutions for the nuclear response have become available during the last decades \cite{Schuck2021}, in the DFT-based calculation schemes a rather good description of the major phonon characteristics for the dynamical kernels can be obtained in QRPA. After formulating our approach in the quasiparticle basis, we notice that the vertex functions $\Gamma^{(11)}$ and $\Gamma^{(02)}$ (\ref{Gamma11_HFB},\ref{Gamma02_HFB}) have the same structure as the variation of the quasiparticle Hamiltonian:
\bea
\delta H^{(11)} = U^{\dagger}\delta h U &+& U^{\dagger}\delta\Delta^{(+)}V \nonumber\\&-& V^{\dagger}\delta\Delta^{(-)\ast}U 
- V^{\dagger}\delta h^T V \\
-\delta H^{(02)} = V^{T}\delta h U &+& V^{T}\delta\Delta^{(+)}V \nonumber\\&-& U^{T}\delta\Delta^{(-)\ast}U 
- U^{T}\delta h^T V,
\label{dH}
\eea
which enter the FAM-QRPA equations  \cite{Bjelcic2020}:

\bea
\delta{\cal R}^{(20)}_{\mu\nu}(\omega)  = \frac{\delta{\cal H}^{(20)}_{\mu\nu}(\omega) + F^{(20)}_{\mu\nu}}{\omega - E_{\mu} - E_{\nu}} \nonumber\\
\delta{\cal R}^{(02)}_{\mu\nu}(\omega)  = \frac{\delta{\cal H}^{(02)}_{\mu\nu}(\omega) + F^{(02)}_{\mu\nu}}{-\omega - E_{\mu} - E_{\nu}}.
\label{FAM}
\eea
The variation of the quasiparticle Hamiltonian has the following component structure in the quasiparticle basis:%in the FAM is determined by the numerical differentiation 
\bea
\delta{\cal H}(\omega) \equiv \left( \begin{array}{cc} \delta{\cal H}^{(11)}(\omega) & \delta{\cal H}^{(20)}(\omega) \\
-\delta{\cal H}^{(02)}(\omega) & -\delta{\cal H}^{(11)T}(\omega)\end{array}\right), %= \nonumber\\
%= \lim\limits_{\eta\to 0}\frac{1}{\eta}\Bigl( {\cal H}({\cal R}_0 + \eta\delta{\cal R}(\omega)) - {\cal H}({\cal R}_0)\Bigr)
\eea
%with ${\cal R}_0$ being the ground state density \cite{Valatin1961,RingSchuck1980}. 
as well as the density variation and the external field operator. Eqs. (\ref{FAM}) can be solved with the aid of the linearization technique \cite{Kortelainen2015,Bjelcic2020}.
The variations $\delta{\cal H}^{(20)}(\omega)$ and  $\delta{\cal H}^{(02)}(\omega)$ depend on the density variations induced by the external field, so that the non-linear Eqs. (\ref{FAM}) should be solved in a self-consistent iterative cycle. The expansion of $\delta{\cal H}^{(20)}(\omega)$ and  $\delta{\cal H}^{(02)}(\omega)$ in
terms of $\delta{\cal R}^{(20)}(\omega)$ and $\delta{\cal R}^{(02)}(\omega)$ up to linear order leads to the conventional QRPA equations, that is sufficient for the PVC vertices, if an effective interaction is used in the calculations. The obvious advantage of the FAM is that it involves only one-body matrix elements, and no two-body matrix elements enter the calculation scheme, in contrast to the standard diagonalization of the QRPA matrix containing the matrix elements of the two-body interaction \cite{RingSchuck1980}. 

In turn, the variations of the single-particle Hamiltonian $\delta h$ and the pairing fields $\delta\Delta^{(\pm)}$, are related to the effective interaction of the DFT,  which plays the role of $\bar v$ in the DFT-based calculations, so one can assume: 
\bea
\delta h_{12}(\omega) &=& \sum\limits_{34} {\bar v}_{1423}\delta\rho_{34}(\omega), \nonumber\\
\delta\Delta^{(\pm)}_{12}(\omega) &=& \frac{1}{2} \sum\limits_{34} {\bar v}_{1234}\delta\varkappa^{(\pm)}_{34}(\omega).
\label{ddens}
\eea
The density variations $\delta\rho(\omega)$ and $\delta\varkappa^{(\pm)}(\omega)$ are obtained from the solutions of the non-homogeneous (FAM)-QRPA equations with the external field as a free term, while the transition densities $\rho^{\mu}$ and $\varkappa^{\mu(\pm)}$ can be extracted from the solutions of the homogeneous QRPA equations, namely the equations 
\be
\left(\begin{array}{cc} A & B \\ B^{\ast} & A^{\ast} \end{array}\right)  \left(\begin{array}{c} X(\omega)\\ Y(\omega)\end{array}\right) +  
 \left(\begin{array}{c} F^{20}\\ F^{02}\end{array}\right) = \omega \left(\begin{array}{c} X(\omega)\\ -Y(\omega)\end{array}\right) %\nonumber\\
\label{QRPAF}
\ee
and
\be
\left(\begin{array}{cc} A & B \\ B^{\ast} & A^{\ast} \end{array}\right)  \left(\begin{array}{c} X^n\\ Y^n\end{array}\right) 
 = \omega_n \left(\begin{array}{c} X^n\\ -Y^n\end{array}\right), \nonumber\\
\label{QRPA}
\ee
respectively. In Eqs. (\ref{QRPAF},\ref{QRPA}), $A$ and $B$ are the regular QRPA matrices \cite{RingSchuck1980}, $\delta{\cal R}^{(20)}(\omega) = X(\omega)$, and $\delta{\cal R}^{(02)}(\omega) = Y(\omega)$. Similarly to the case of the quasiparticle propagator, for both  the density variations and the transition densities their components in the canonical basis are mapped to the $X$ and $Y$ components in the quasiparticle basis:
\bea
\delta\rho_{12}(\omega) &=&  (UX(\omega)V^T + V^{\ast}Y^T(\omega)U^{\dagger})_{12} \nonumber\\
\delta\varkappa^{(+)}_{12}(\omega) &=& (UX(\omega)U^T + V^{\ast}Y^T(\omega)V^{\dagger})_{12}\nonumber\\
\delta\varkappa^{(-)}_{12}(\omega) &=& (V^{\ast}X^{\dagger}(\omega)V^{\dagger} + UY^{\ast}(\omega)U^T)_{12}
\eea
and
\bea
\rho^n_{12} &=&  (UX^nV^T + V^{\ast}Y^{nT}U^{\dagger})_{12} \nonumber\\
%
%\rho^{n\dagger}_{12} =  (V^{\ast}X^{n\dagger}U^{\dagger} + UY^{n\ast}V^T)_{12} \ \ \ \ \ \ \  \delta\rho^{\dagger}_{12}(\omega) =  (V^{\ast}X^{\dagger}
%(\omega)U^{\dagger} + UY^{\ast}(\omega)V^T)_{12}\nonumber\\
%
\varkappa^{n(+)}_{12} &=& (UX^nU^T + V^{\ast}Y^{nT}V^{\dagger})_{12} \nonumber\\ 
\varkappa^{n(-)}_{12} &=& (V^{\ast}X^{n\dagger}V^{\dagger} + UY^{n\ast}U^T)_{12},
\eea
respectively.
Therefore, their components in the canonical basis are related at the poles of the QRPA propagator $\omega_{\mu}$ as follows \cite{Hinohara2013,Litvinova2021}:
\bea
\delta\rho_{12}(\omega\to \omega_{\mu}) &=&
\frac{\rho^{\mu}_{12}\langle {\mu}|F|0\rangle }{\omega - \omega_{\mu} + i\delta}\nonumber\\
\delta\varkappa^{(+)}_{12}(\omega\to \omega_{\mu}) &=&  
\frac{\varkappa^{\mu(+)}_{12}\langle {\mu}|F|0\rangle }{\omega - \omega_{\mu} + i\delta}\nonumber\\
\delta\varkappa^{(-)\ast}_{12}(\omega\to \omega_{\mu}) &=&  
\frac{\varkappa^{\mu(-)\ast}_{12}\langle {\mu}|F|0\rangle }{\omega - \omega_{\mu} + i\delta}.
\label{densdens}
\eea
With the aid of Eqs. (\ref{Gamma11_HFB} -- \ref{densdens}), one can see that the qPVC vertices $\Gamma^{(11)\mu}$ and $\Gamma^{(02)\mu}$ 
and the variations of the quasiparticle Hamiltonian $\delta H^{(11)}$ and $\delta H^{(02)}$ at the peaks of the strength function $\omega = \omega_{\mu}$ are related by
\be
\Gamma^{(ij)\varkappa}_{\nu\nu'} = \lim\limits_{\delta\to 0}\sqrt\frac{\delta}{\pi S(\omega_{\varkappa})}\text{Im}\Bigl(\delta{\cal H}^{(ij)}_{\nu\nu'}(\omega_{\varkappa}+i\delta)\Bigr),
\label{PVCvertex}
\ee
up to an unimportant phase. The values of the strength function $S(\omega)$ 
\be
S(\omega) = -\frac{1}{2\pi} \text{Im}\sum\limits_{\mu\nu}\Bigl(F^{(20)\ast}_{\mu\nu}\delta{\cal R}^{(20)}_{\mu\nu}(\omega) + F^{(02)\ast}_{\mu\nu}\delta{\cal R}^{(02)}_{\mu\nu}(\omega)\Bigr)
\ee
at the peaks of its distribution enter Eq. (\ref{PVCvertex}) to correctly normalize the vertices by removing the dependence on the external field, that becomes exact at $\delta\to 0$. Alternatively, the vertices can be extracted by contour integrations of the density or quasiparticle Hamiltonian variations in the complex plane, as described in Refs. \cite{Hinohara2013,Litvinova2021}.  

\subsection{Sum rules for quasiparticle states}

The energies of fragmented states ${\cal E}_n$ and the corresponding spectroscopic factors $S^{(\eta)n}_{\nu\nu'}$ entering the correlated propagators in Eq. (\ref{G_qp})  satisfy the sum rules, which relate these quantities to their mean-field (RHB) counterparts. Such sum rules for the non-superfluid case appear, for instance, in Refs. \cite{Baranger1970,Birbrair2000} in the context of the Baranger theorem, while below we formulate them for the superfluid case.

The correlated quasiparticle propagator (\ref{G_qp}) can be expanded in negative powers of the energy variable in the high-energy limit $\varepsilon \to \infty$ as:
\be
G^{(\eta)}_{\nu\nu'}(\varepsilon) = \frac{I^{(0)}_{\nu\nu'}}{\varepsilon} + \frac{I^{(1)}_{\nu\nu'}}{\varepsilon^2} + \frac{I^{(2)}_{\nu\nu'}}{\varepsilon^3} + ...,
\label{G_asymp}
\ee
where
\bea
I^{(0)}_{\nu\nu'} &=& \sum\limits_n S^{(\eta)n}_{\nu\nu'} \nonumber\\ 
I^{(1)}_{\nu\nu'} &=& \sum\limits_n \eta {\cal E}_n S^{(\eta)n}_{\nu\nu'}\nonumber\\
I^{(2)}_{\nu\nu'} &=& \sum\limits_n (\eta {\cal E}_n)^2 S^{(\eta)n}_{\nu\nu'},
\label{I0}
\eea
and we set $E_0 = 0$.  Eq. (\ref{I0}) can be verified with the aid of the geometrical progression summation formula for $\varepsilon > E$:
\be
\frac{S}{\varepsilon - E} = \frac{S}{\varepsilon(1 - E/\varepsilon)} = \frac{S}{\varepsilon}\sum\limits_{k = 0}^{\infty}\Bigl(\frac{E}{\varepsilon}\Bigr)^k.
\ee
The self-energy (\ref{SEqp}) can be similarly decomposed as
\be
\Sigma^{e(\eta)}_{\nu\nu'}(\varepsilon) = \frac{\Pi^{(0)}_{\nu\nu'}}{\varepsilon} + \frac{\Pi^{(1)}_{\nu\nu'}}{\varepsilon^2} + \frac{\Pi^{(2)}_{\nu\nu'}}{\varepsilon^3} + ...,
\label{SE_asymp}
\ee 
with the obvious meaning of the numerators $\Pi^{(k)}_{\nu\nu'}$. 
Substitution of Eqs.  (\ref{G_asymp},\ref{SE_asymp}) to Eq. (\ref{Dyson_qp}) leads to:
\bea
\varepsilon\Bigl( \frac{I^{(0)}_{\nu\nu'}}{\varepsilon} + \frac{I^{(1)}_{\nu\nu'}}{\varepsilon^2} + \frac{I^{(2)}_{\nu\nu'}}{\varepsilon^3} + ... \Bigr) = 
\delta_{\nu\nu'} \nonumber\\
+ \sum\limits_{\nu''} \Bigl( \delta_{\nu\nu''}\eta E_{\nu} + \frac{\Pi^{(0)}_{\nu\nu''}}{\varepsilon} + \frac{\Pi^{(1)}_{\nu\nu''}}{\varepsilon^2} + \frac{\Pi^{(2)}_{\nu\nu''}}{\varepsilon^3} + ... \Bigr)
\nonumber\\
\times\Bigl( \frac{I^{(0)}_{\nu''\nu'}}{\varepsilon} + \frac{I^{(1)}_{\nu''\nu'}}{\varepsilon^2} + \frac{I^{(2)}_{\nu''\nu'}}{\varepsilon^3} + ... \Bigr).
\eea
Equating the coefficients at the zeroth and negative-one powers of $\varepsilon$, one obtains the non-energy-weighted and the energy-weighted  sum rules, respectively:
\be
\sum\limits_{n} S^{(\eta)n}_{\nu\nu'} = \delta_{\nu\nu'} \ \ \ \ \ \ \ \  \sum\limits_{n} {\cal E}_nS^{(\eta)n}_{\nu\nu'} = \delta_{\nu\nu'}E_{\nu}. 
\label{SumRules1}
\ee  
The non-energy-weighted sum rule reflects the conservation of probabilities for the given quasiparticle state: the occupancy of an HFB state is equal to unity in the HFB basis (being, however, fractional in the canonical basis), that is expressed by $\delta_{\nu\nu'}$ in the numerator of the mean-field quasiparticle propagator $\tilde G$ of Eq. (\ref{Gq}), and this occupancy is equal to the sum of the occupancies for the fragments $S^{(\eta)n}_{\nu\nu'}$ of the given quasiparticle state, when the dynamical qPVC self-energy (\ref{SEqp}) is taken into account. The energy-weighted sum rule expresses the fact that the centroid of the fragmented state is located exactly at the energy of the reference mean-field state. 
%This does not contradict the picture presented in Fig. \ref{250cf}: while we plot only the dominant states with the largest spectroscopic factors in Fig. \ref{250cf},  
%the full solution of Eq. (\ref{Dyson_qp}) returns numerous fragments, most of them with small spectroscopic factors in a large energy interval, that compensates the %relative large shift of the dominant state. 
The sum rules (\ref{SumRules1}) are very useful to control numerical implementations of the approaches which include (q)PVC or singular self-energies of perturbative character. 

We emphasize here that the sum rules (\ref{SumRules1}) are obtained using the definition of the propagators (\ref{Gq}) and  the  form of the self-energy (\ref{SEqp}), which consist of simple poles with the properly normalized residues (notice that this property and, thus, the obtained sum rules are valid also for the exact self-energy with the correlated three-fermion propagators). The latter is a manifestation of locality and unitarity, the typical quantum field theory constraints which 
%translate to 
are compatible with causality in the time domain. Another condition is the full solution of the Dyson equation (not a perturbative expansion), which is also part of the derivation of the sum rules (\ref{SumRules1}).  

We notice also that the completeness of the phonon space and the approximations, in which the phonons are computed, do not play a role for the sum rules (\ref{SumRules1}), which do not even include explicitly the residues of the dynamical self-energy and its poles (these quantities start to appear in higher-power sum rules). The sum rules (\ref{SumRules1}) remain fulfilled for any number of phonons and for any number of intermediate states in Eq. (\ref{SEqp}). The reader can easily verify this statement for the case of one phonon mode and one intermediate quasiparticle state in Eq. (\ref{SEqp}): in this case the secular equation corresponding to Eq. (\ref{Dyson_qp}) reduces to a quadratic equation. Thus, we summarize that truncations of the (q)PVC model space do not violate the sum rules (\ref{SumRules1}).

%%%%%%% Beginning of Letter
%
%===============================================================================
% Calculation details
%===============================================================================
\section{Calculation details, results and discussion}
The numerical implementation of the approach described above is based on the FAM-QRPA of Ref. \cite{Bjelcic2020}, which is employed to generate the quasiparticle-phonon model space in axial geometry. The relativistic Hartree-Bogoliubov (RHB) equations for the stationary fermionic basis states resulting from the relativistic point coupling Lagrangian
were solved by expanding the Dirac spinors in terms of eigenfunctions of an axially symmetric harmonic oscillator potential. Ten major shells were used in the calculations for light nuclei with the mass numbers $A \approx 30-40$ and the number of the oscillator shells was extended to fourteen in the calculations for heavy nuclei with masses around $A \approx 250$. The density-dependent point-coupling interaction DD-PC1 \cite{DD-PC1} and the finite-range pairing force with D1S parametrization \cite{D1S} in the separable form \cite{Tian2009,Tian2009a} were employed in the calculations. The FAM-QRPA equations (\ref{FAM}) were solved iteratively with the aid of the modified Broyden's method \cite{Baran2008} and with the convergence criteria defined in Ref. \cite{Bjelcic2020}. The imaginary part $\delta = 100$ keV of the frequency argument $\omega$ was employed to eliminate the divergencies of the subsequently computed strength distribution $S(\omega)$ at the roots of Eqs. (\ref{FAM}). This value of $\delta$ is sufficiently small for the extraction of the qPVC vertices by Eq. (\ref{PVCvertex}) with a reasonable accuracy. Both normal and pairing phonon modes with $J^{\pi}$ = 2$^+$, 3$^-$, 4$^+$, 5$^-$ and $0\leq K\leq J$ were included in the quasiparticle dynamical self-energy (\ref{SEqp}). Although it is technically difficult to extend the calculations beyond $J = 5$ at this point, we have found gradually decreasing contributions from large-J phonons, similarly to the spherical case. Contributions from the $J^{\pi} $= 0$^+$ and $J^{\pi}$ = 1$^-$
 were found negligible. The dynamical self-energy (\ref{SEqp}) was treated in the diagonal approximation $|\nu\rangle = |\nu'\rangle$, which was found quite accurate in the calculations for spherical nuclei \cite{LitvinovaRing2006,LitvinovaAfanasjev2011,AfanasjevLitvinova2015,Litvinova2012,Litvinova2016}. It is expected to be a  good approximation also for deformed systems because of destructive interference between the non-diagonal terms.
% justified by the angular momentum conservation at the qPVC vertices. 
The phonon frequency cutoff $\omega_{\varkappa}^{max} = 15$ MeV was adopted for $\omega_{\varkappa}$. The phonon modes within each $\{J^{\pi}, K\}$ family were selected by their reduced transition probabilities of the electric multipole transitions: the phonons with the reduced transition probabilities equal or exceeding 10\% of the maximal one were kept in the model space. The 
quasiparticle intermediate states $|\nu''\rangle$ with the energy differences  $|E_{\nu}- E_{\nu''}| \leq$ 60 MeV were included in the summation of Eq. (\ref{SEqp}), that ensured its convergence. This calculation scheme allowed us to include the leading contributions to Eq. (\ref{SEqp}) and it is justified by the preceding qPVC calculations for medium-heavy spherical nuclei \cite{LitvinovaAfanasjev2011,AfanasjevLitvinova2015,Litvinova2012}. 

\begin{figure*}
\begin{center}
\vspace{-0.5cm}
\includegraphics[scale=0.50]{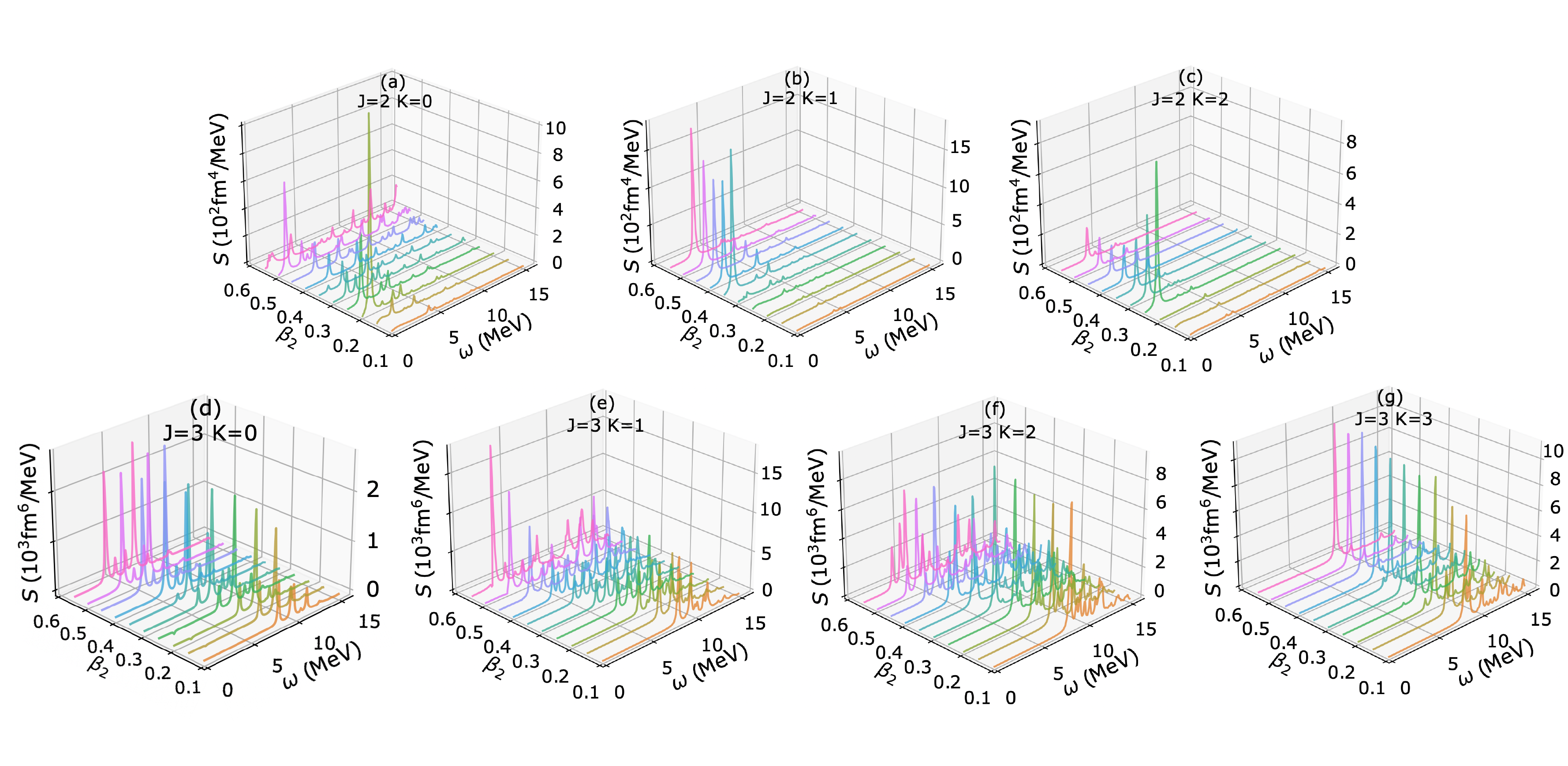}
\end{center}
\vspace{-0.5cm}
\caption{The  $J^{\pi} = 2^+$ and $J^{\pi} = 3^-$ low-energy isoscalar strength functions for varying quadrupole deformation in $^{38}$Si computed within the relativistic FAM-QRPA of Ref. \cite{Bjelcic2020}. The smearing parameter $\delta$ = 100 keV (imaginary part of the energy argument) was used in the calculations.  }
\label{38si_Strength}
\end{figure*}

Fig. \ref{38si_Strength} displays the FAM-QRPA responses to isoscalar operators with  $J^{\pi} = 2^+$ and $J^{\pi} = 3^-$ in $^{38}$Si, illustrating their evolutions with quadrupole deformation parameter $\beta_2$. In the approaches based on effective nucleon-nucleon interactions, as in this work, QRPA  provides a reasonable description of both low-energy and high-energy collective states. Although the observed response indicates that correlations of higher complexity than those of QRPA are needed to describe the excitation spectra \cite{LitvinovaRingTselyaev2008,LitvinovaRingTselyaev2010,LitvinovaSchuck2019}, QRPA phonons are sufficient to capture the leading qPVC effects in both the one-fermion and two-fermion self-energies. This point was investigated and confirmed explicitly in Ref. \cite{Tselyaev2018} in beyond-QRPA calculations based on the Skyrme EDF.
%while in general it is in compliance with the Migdal theorem stating that the PVC-type corrections to fermionic lines quantitatively dominate the corrections to the PVC vertices \cite{FetterWalecka}. 
In ab-initio frameworks based on the bare nucleon-nucleon interaction (Q)RPA, however, produces too unrealistic results for the nuclear response and, thus, for the phonon modes, so that higher-complexity approaches beyond (Q)RPA should be employed  \cite{PapakonstantinouRoth2009,Bianco2012,DeGregorio2016a}.  Fully ab-initio solutions for fermionic EOMs, such as the one described here and  in Refs. \cite{LitvinovaSchuck2019,LitvinovaSchuck2020}, remain  tasks for future research. 

After obtaining the spectra of quasiparticles and phonons, the matrix elements $\delta{\cal H}^{(11)}$ and $\delta{\cal H}^{(02)}$ were retrieved at the energies corresponding to the roots of Eq. (\ref{FAM}). Subsequently, the qPVC vertices were extracted with the aid of Eq. (\ref{PVCvertex}) for the selected phonon modes.
This information was then used for constructing the dynamical self-energy of Eq. (\ref{SEqp}). With this input, the Dyson equation (\ref{Dyson_qp}) was transformed to the arrowhead matrix form as in Refs. \cite{RingWerner1973,LitvinovaRing2006} and solved by the ordinary diagonalization procedure. In this work we focused on the quasiparticle states states located within $E_w \approx\pm 10$ MeV energy window around the Fermi energy, and Eq. (\ref{Dyson_qp}) was solved separately for each of these states. The spectroscopic factors were determined via the derivatives of the dynamical self-energy at the poles of the resulting quasiparticle propagator
\be
S^{(\eta)n}_{\nu\nu'} = \bigl[ \delta_{\nu\nu'} - \frac{d\Sigma^{r(\eta)}_{\nu\nu'}(\varepsilon)}{d\varepsilon}\vert_{\varepsilon=\eta {\cal E}_n}\Bigr]^{-1},
\ee
as it follows from Eqs. (\ref{Dyson_qp},\ref{G_qp}).

%
%===============================================================================
% Results
%===============================================================================
 %
\begin{figure}
\begin{center}
\vspace{-0.5cm}
\includegraphics*[scale=0.60]{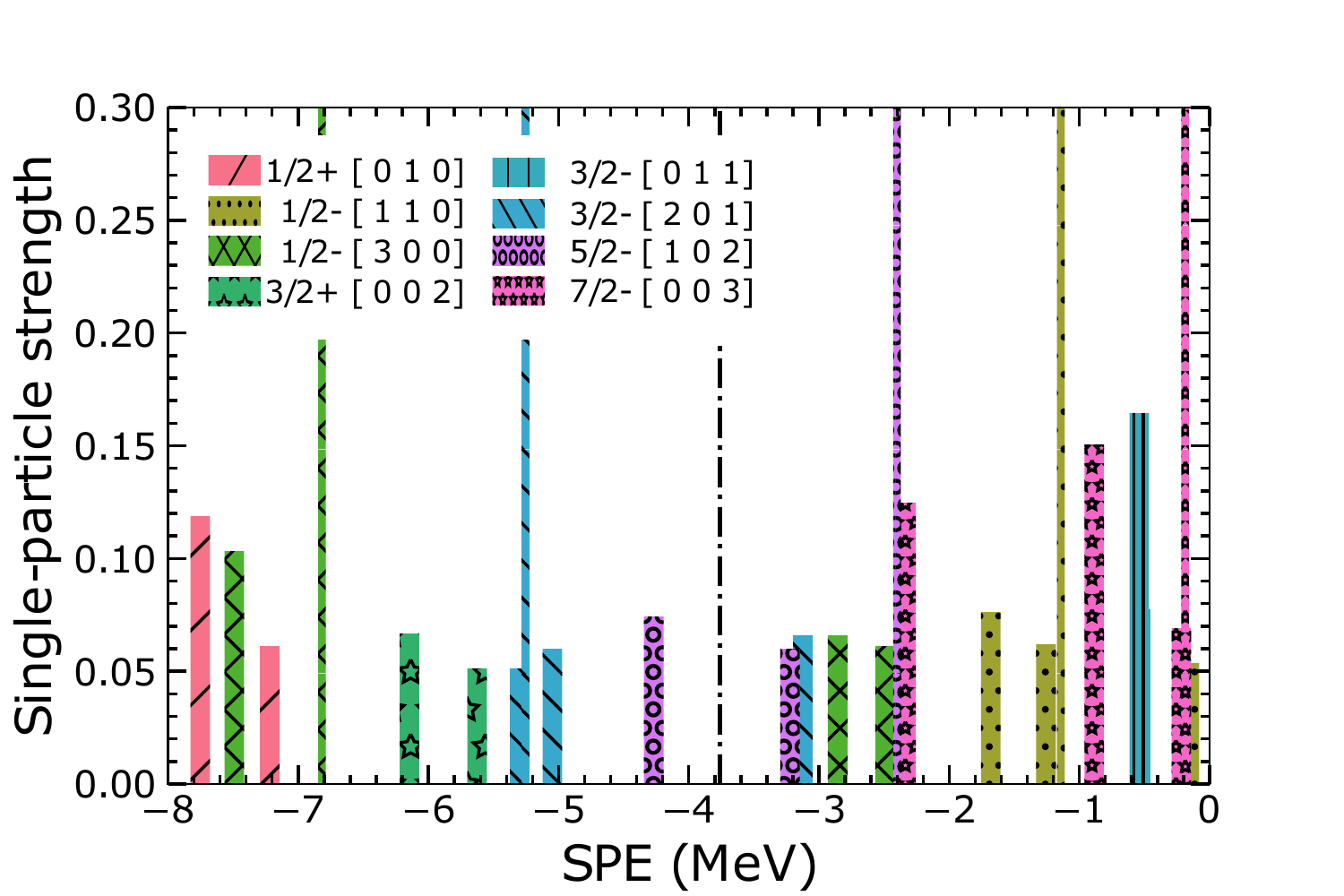}
\end{center}
\vspace{-0.5cm}
\caption{Single-particle energies (SPE) and strength of the neutron states in the axially deformed $^{38}$Si obtained in the RHB-PVC calculations (thick bars) in comparison with the RHB reference states (thin bars). The vertical dash-dotted line denotes the RHB Fermi energy. }
\label{si_neutrons}%
\end{figure}
%

%{\it Results. \textemdash}
The calculations were performed for a set of light and heavy nuclei with axial deformations. Fig. \ref{si_neutrons} displays the correlated neutron quasiparticle states obtained in the RHB-PVC  calculations for the neutron-rich nucleus $^{38}$Si with the self-consistent prolate deformation $\beta_2$ = 0.31. The thick bars represent the fragments of the final quasiparticle states located at $\lambda\pm {\cal E}_n$, i.e., above (below) the RHB Fermi energy $\lambda$ if their RHB occupancies are smaller (greater) than 0.5. Their heights correspond to the spectroscopic factors of these states in the canonical basis. The RHB reference states at energies $\lambda\pm E_{\nu}$ are given by the thin bars with the unity hight. The comparison between the thin and thick bars reveals the effects of the qPVC in the nucleonic dynamical self-energy on the quasiparticles. One can see that a remarkable fragmentation occurs already at the Fermi surface indicated by the dash-dotted line. The analytic structure of the dynamical self-energy $\Sigma^{r(\eta)}_{\nu\nu'}(\varepsilon)$ implies that each RHB basis state $|\nu\rangle$ splits into a large number of fragments corresponding to the number of terms in Eq. (\ref{SEqp}).
%, that is a product of the number of the intermediate quasiparticle states $|\nu''\rangle$ and the phonon modes $\varkappa$. 
The first, forward-going, term is responsible for the main qPVC effect and the second, backward going one, is the counter term famously associated with the ground state correlations, which reduce the qPVC. As a result, the major part of the obtained correlated quasiparticle states are represented by a few competing fragments as, for instance,  the states $1/2^+[010], 1/2^-[110], 3/2^+[002]$ and $3/2^-[201]$.
%and $5/2^-[102]$  below the Fermi energy, and the states $1/2^-[011]$ and $3/2^+[002]$ above it. 
These states are characterized by the presence of two or three fragments with comparable spectroscopic factors $S_{\nu\nu}^{(\pm)n}$ of the order of 0.1-0.2 units. This is a new feature as compared to the previously studied spherical nuclei, where typically a dominant fragment with large spectroscopic factor can be extracted for the states at the Fermi energy, with the most pronounced dominance in closed-shell systems. The axial deformation, together with the superfluid pairing correlations in deformed open-shell nuclei, induce a considerably stronger fragmentation, which can be linked to the fact that these two effects stipulate the formation of the collective phonon modes at lower energies. The overall trend is, however, similar to that found for spherical nuclei: the center of gravity of the major fragments is moving toward the Fermi energy, with respect to the reference RHB quasiparticle states.

%"Non-observable character of the spectroscopic factors"... ?
%Another group of quasiparticle orbits: $5/2^-[102], 1/2^-[110]$ and $1/2^-[011]$  above the Fermi energy show fragmentation with the dominance of one fragment %located below the RHB reference state. 
 %
\begin{figure}
\begin{center}
\vspace{-1.5cm}
\includegraphics[scale=0.62]{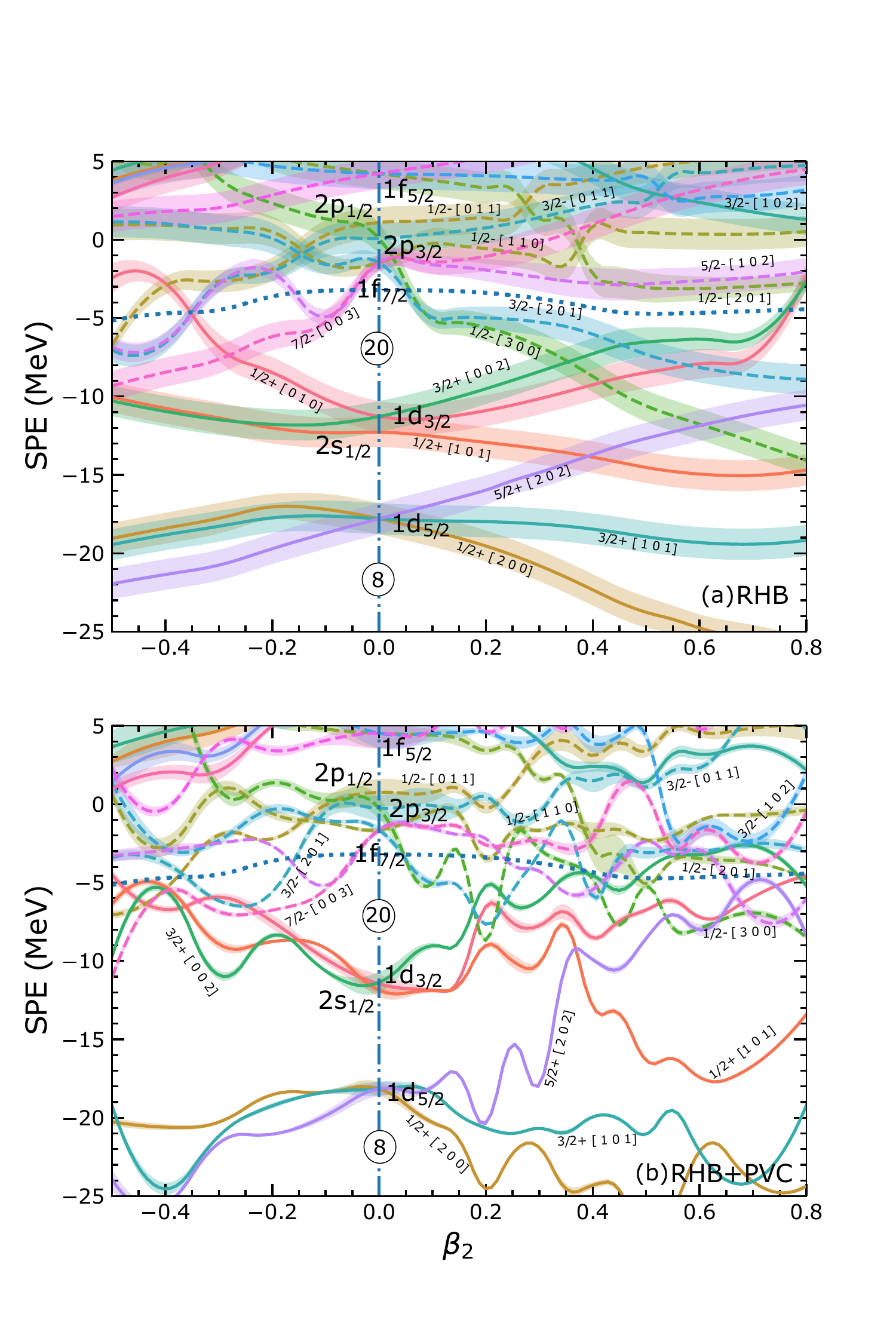}
\end{center}
\vspace{-1.3cm}
\caption{The Nilsson diagram for $^{38}$Si extracted from the RHB calculations (top) and from the RHB-PVC calculations (bottom).}
\label{nilsson}%
\end{figure}
In order to illuminate the effect of deformation, we performed similar calculations with different values of the deformation parameter $-0.5\leq \beta_2\leq 0.8$ spanning a wide range from prolate to oblate deformations with the step of 0.05. The results for the dominant fragments, i.e, the fragments with the largest spectroscopic factors 
%$S^{(\eta)n_0}_{\nu\nu}$, 
and with the energies ${\cal E}_{n_0}$ of the correlated neutron quasiparticle states are displayed in the bottom panel of Fig. \ref{nilsson}. Their energies are plotted as functions of the deformation parameter $\beta_2$ and compared to the RHB Nilsson diagram shown in the top panel. The finite width of each color band is proportional to the value of 
%%%%
$v^2_{\nu}S^{(\eta)n_0}_{\nu\nu}$ and $(1-v^2_{\nu})S^{(\eta)n_0}_{\nu\nu}$, with $v^2_{\nu}$ being the RHB occupancies, for the states below and above the Fermi energy, respectively, as these products represent the resulting single-particle spectroscopic factors. 
%%%
The noticeably thinner bands in the case of the RHB-PVC states indicate the considerable reduction of the occupancies with respect to the pure RHB calculations, if only one dominant fragment is taken into account.

However, Fig. \ref{si_neutrons} discussed above is complementary to Fig. \ref{nilsson} as well as the non-energy-weighted and energy-weighted sum rules (\ref{SumRules1}),
which reflect the conservation of probabilities (spectroscopic factors) and centroids of the correlated quasiparticle states. 
%The former sum rule expresses the fact that the spectroscopic factors of the fragments for each RHB state $|\nu\rangle = |\nu'\rangle$ add up to one and %the latter implies the energy centroid conservation, if the complete set of fragments is included in the sums (\ref{SumRules}).
Both sum rules are fulfilled in our numerical implementation with high accuracy. 
One can see in Fig. \ref{nilsson} that both the energies and the occupancies of the dominant fragments show notable variations with the deformation parameter. First of all, we emphasize that for the vanishing deformation parameter the calculations in the axial symmetry yield the correct limit, which is verified by the degeneracy of the quasiparticle states at $\beta_2=0$ reproduced with good accuracy.  The occupancies of the dominant fragments are maximized at the spherical shape. The next observation is the additional oscillations of the positions of the dominant fragments on the energy scale with respect to the relatively smooth evolution of the RHB states with the deformation. Such oscillations are attributed to the evolution of the low-energy collective phonon modes, which play the major role in the qPVC, with deformations. The corresponding isoscalar strength functions for 
$J^{\pi}$ = 2$^+$ and 3$^-$ in $^{38}$Si shown in Fig. \ref{38si_Strength} illustrate this evolution.
We observe, for instance, the disappearance of the $J = 2$  low-lying states with $K = 0$ and the simultaneous appearance of the $J = 2$ and $J = 3$  low-energy modes with $K = 1$ as well as the $J = 2, K = 2$ one in the interval $\approx0.3\leq\beta_2\leq 0.6$, while the $J = 2, K = 0$ mode reappears again at $\beta_2 = 5.5$. Similar irregularities are observed in $J = 4$ and $J = 5$ channels. The non-smooth behavior of quite a few dominant quasiparticle states in this interval is a direct consequence of these irregularities in the phonon spectra. Remarkably, this effect gives rise to the formation of the new shell closure with the neutron number $N = 12$ at  $\beta_2 \geq 0.5$.

\begin{figure}
\begin{center}
%\vspace{-0.3cm}
\includegraphics*[scale=0.37]{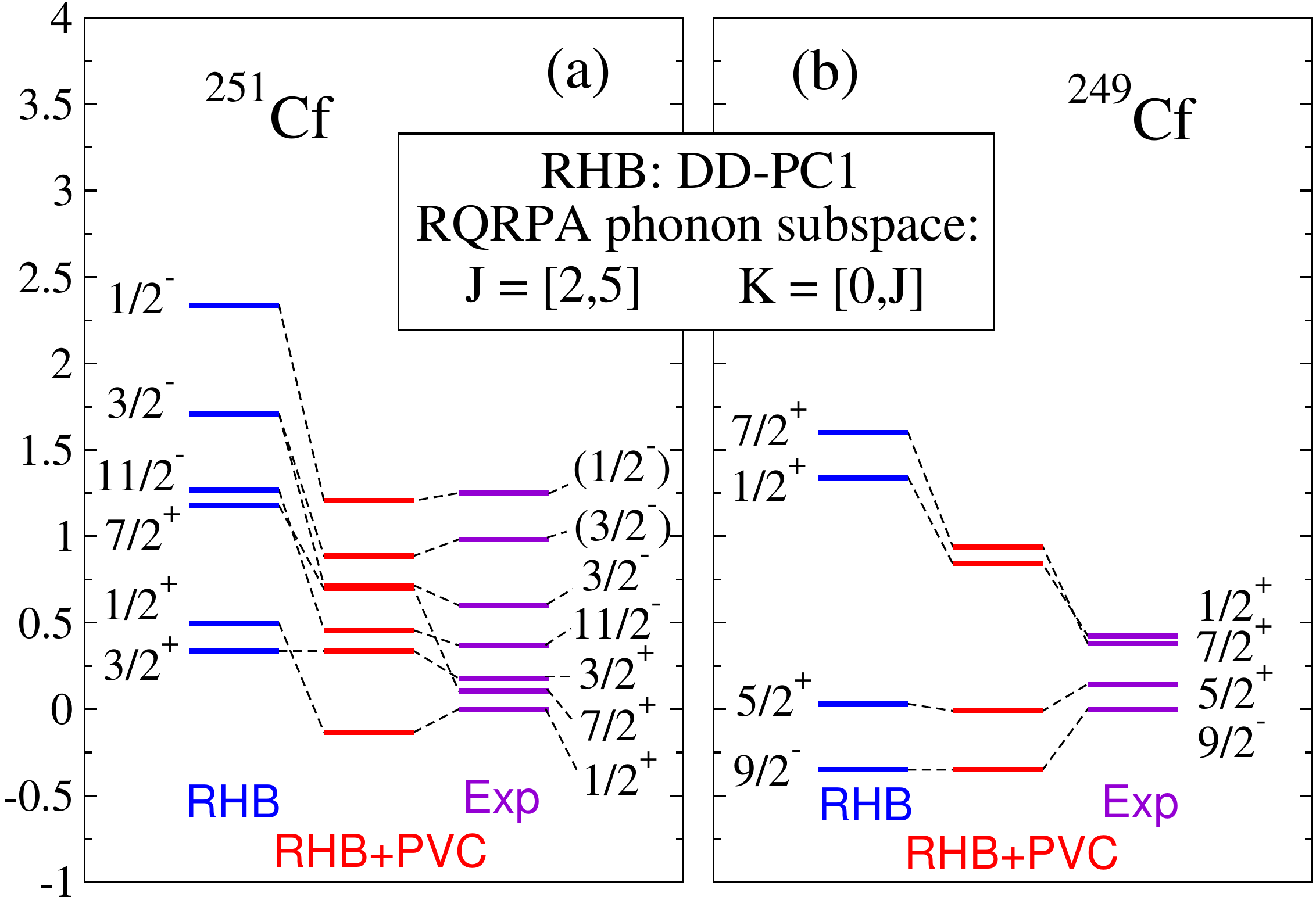}
\end{center}
\vspace{-0.3cm}
\caption{Single-quasiparticle neutron states in the axially deformed $^{251}$Cf (a) and $^{249}$Cf (b). Left columns: the RHB states, middle columns: the RHB-PVC dominant states, right columns: experimental data from Ref. \cite{NNDC}. }
\label{250cf}%
\end{figure}

Fig. \ref{250cf} represents our results  for the single-quasiparticle states in heavy nuclei displaying the level scheme  for the neutron subsystems of $^{251,249}$Cf, whose experimental ground states are taken as a  reference at $E = 0$. The RHB ground states of these nuclei are supposed to be unaffected by PVC as the parameters of the DFT are adjusted to nuclear ground states. 
The parameter of axial deformation $\beta_2 = 0.29$ was obtained in the self-consistent RHB calculations for $^{250}$Cf, in agreement with its experimental value \cite{NNDC}. 
%The relativistic FAM-QRPA calculations reveal a number of low-lying phonon modes {\bf in  $J^{\pi}$ = 2$^+$, 3$^-$, 4$^+$, 5$^-$, $0\leq K\leq J$ %channels in the $^{250}$Cf core, which are known to be present in its experimental spectrum  \cite{NNDC} and couple to the RHB quasiparticle states %with considerable strength. 
 The dominant quasiparticle states, extracted from the solution of Eq. (\ref{Dyson_qp}) for the poles and residues of the quasiparticle propagator $G^{(\eta)}_{\nu\nu'}(\varepsilon)$, above and below the Fermi surface are identified with those in the neighboring $N \pm 1$ nuclei, respectively, according to the definition of those poles 
 %Assuming that the dominant fragments are more likely to be detected in the experimental probes of the single-quasiparticle states, we omit the fragments other than the dominant ones in Fig \ref{250cf}, in order to 
 and compared to the available data on the band-head levels in $^{251}$Cf and $^{249}$Cf from the data base \cite{NNDC}. As it can be seen from Fig. \ref{250cf}, qPVC causes sizable shifts of the energies of the dominant fragments obtained in the RHB-PVC calculations with respect to the reference RHB  states. All the shifts are directed downward leading to the overall compression of the single-quasiparticle spectra, 
while for the majority of the levels the RHB+PVC results are of nearly spectroscopic accuracy. Remarkably, the shift and splitting of the 3/2$^-$ RHB state is accurately reproduced, and the ground state spin of $^{251}$Cf is changed by qPVC, in agreement with data.
An additional enhancement of the level density occurs because of the strong fragmentation of the states, similar to the case of $^{38}$Si and in agreement with the phenomenological model of Ref. \cite{Malov1976}.

To overcome the remaining minor discrepancies between theory and experiment, the approach can be further perfected by (i) relaxing the diagonal approximation for the self-energy (\ref{SEqp}), (ii) including the phonon modes with unnatural parities and isospin flip, which are known to make generally a weaker contribution than the neutral natural parity phonons, but cumulatively may slightly further reinforce the qPVC effects, (iii) elaborating on a subtraction procedure for the nucleonic self-energy to remove the double counting of qPVC, which should be removed in the DFT-based implementations, when the qPVC model space is close to completeness. Such a procedure has become a common practice in DFT-based applications for nuclear response \cite{LitvinovaRingTselyaev2008,LitvinovaRingTselyaev2010,Gambacurta2015,Tselyaev2018,Lyutorovich:2018cph,LitvinovaSchuck2019}, being proposed originally in Ref. \cite{Tselyaev2013}. For the case of the single-quasiparticle EOM such a procedure has not been developed yet, however, the first steps toward its understanding are made in Refs. \cite{LitvinovaSchuck2019,Litvinova2021}, where the single-(quasi)particle EOM is derived in the ab-initio framework. 
%, whose positions were corrected by the respective binding energy differences in $^{250}$Cf and the neighboring odd nuclei. Thereby, one can see that the calculated energies of the dominant fragments move toward the experimental ones, that improves the picture considerably. This effect stems solely from the qPVC and confirms that the inclusion of this mechanism is necessary for describing the quasiparticle states in deformed nuclei.

%===============================================================================
% Summary
%===============================================================================
%{\it Summary. \textemdash}
\section{Summary}
We presented a framework which allows for a synthesis of the two powerful techniques: the equation of motion 
for the fermionic correlation functions and the finite amplitude method for vibrational modes in nuclei. 
%%%%  Added to regular article %%%%
The EOM for the quasiparticle propagator in a superfluid medium obtained from the bare fermionic Hamiltonian in the form of Dyson equation contains static and dynamical interaction kernels in the most general exact form. The three-fermion propagators of the dynamical kernel can be with a good accuracy factorized into two-fermion and one-fermion ones which, in the superfluid case, generates the coupling of quasiparticles to superfluid phonons. The latter phonons unify the normal and pairing phonons and, in general, are the solutions of the two-quasiparticle EOM. In this work we approximated the superfluid phonons by the relativistic QRPA, that is compatible with the static kernel of the quasiparticle EOM in the form of the relativistic Hartree-Bogoliubov approach. Furthermore, the link between the two EOMs allows for establishing a relationship between the qPVC vertices in the dynamical kernel and the variations of the RHB Hamiltonian in FAM-QRPA.  

%%%%

The latter provides efficient computation
of the phonon frequencies and the quasiparticle-vibration coupling vertices, which are incorporated into the Dyson equation for the nucleonic  propagator. The approach is formulated in the 
basis of Dirac-Hartree-Bogoliubov quasiparticles and implemented for open-shell nuclei with axial deformations. The analysis of the solutions obtained  for the medium-light neutron-rich nucleus $^{38}$Si and for the heavy nucleus $^{250}$Cf reveals a significant fragmentation of the quasiparticle states around the Fermi surfaces and an increase of the level densities in both neutron and proton subsystems. This improves considerably the agreement with experimental data for axially-deformed nuclei as compared to the mean-field approximation. The developed framework and its numerical implementation open the way for further progress on computation of the nuclear spectral properties in non-spherical geometries.

%===============================================================================
% Acknowledgements
%===============================================================================
%{\it Acknowledgements. \textemdash}
\section{Acknowledgements}
Fruitful discussions with Anatoli Afanasjev are gratefully acknowledged.
This work is supported in parts by the US-NSF Career Grant PHY-1654379 and by the QuantiXLie Centre of Excellence, a project co-financed by the Croatian Government and European Union through the European Regional Development Fund - the Competitiveness and Cohesion Operational Programme (KK.01.1.1.01.0004).
%
%===============================================================================

%\widetext
%\appendix

%\vspace*{0.5cm}
%\newpage

\bibliography{Bibliography_Mar2021}

\end{document}